\documentclass[11pt]{article}
\usepackage{Blois,epsfig}

\def\be{\begin{equation}}
\def\ee{\end{equation}}
\def\bea{\begin{eqnarray}}
\def\eea{\end{eqnarray}}

\begin{document}
\vspace*{2cm}
\begin{center}
\Large{\textbf{XIth International Conference on\\ Elastic and Diffractive Scattering\\ Ch\^{a}teau de Blois, France, May 15 - 20, 2005}}
\end{center}

\vspace*{2cm}
\title{Hard Diffraction --- from Blois 1985 to 2005}

\author{Gunnar Ingelman}

\address{High Energy Physics, Uppsala University, Box 535, SE-75121 Uppsala, Sweden}

\maketitle\abstracts{The idea of diffractive processes with a hard scale involved, to resolve the underlying parton dynamics, was presented at the first Blois conference in 1985 and experimentally verified a few years later. Today hard diffraction is an active research field with high-quality data and new theoretical models. The trend from Regge-based pomeron models to QCD-based parton level models has given insights on QCD dynamics involving perturbative gluon exchange mechanisms, including the predicted BFKL-dynamics, as well as novel ideas on non-perturbative colour fields and their interactions.}

\section{Idea and discovery of hard diffraction}\label{sec:idea}
At the first Blois meeting in 1985 we presented \cite{Ingelman:1985ec} our novel idea \cite{Ingelman:1984ns} to consider a hard scale in diffractive scattering to resolve an underlying parton level interaction and thereby be able to investigate the process in a modern QCD-based framework. We formulated this in a model with an effective pomeron flux in the proton, $f_{I\!\!P /p}(x_{I\!\!P},t)$, and parton distributions in the pomeron $f_{q,g/I\!\!P}(z,Q^2)$, such that cross-sections for hard diffractive processes could be calculated from the convolution,  
$d\sigma \sim f_{I\!\!P /p} \; f_{q,g/I\!\!P} \; f_{q,g/p} \; d\hat{\sigma}_{\rm{pert.\ QCD}}$, 
of these functions with a perturbative QCD cross-section for a hard parton level process. This enabled predictions of diffractive jet production at the CERN $p\bar{p}$ collider and also of diffractive deep inelastic scattering. Although this seems quite natural in today's QCD language, it was rather controversial at the time. 

It was therefore an important break-through when the UA8 experiment at the CERN $p\bar{p}$ collider actually discovered \cite{Bonino:1988ae} hard diffraction by triggering on a leading proton in their Roman pot detectors and finding jets in the UA2 central calorimeter in basic agreement with our model. The observed jets had normal jet properties and by investigating their longitudinal momentum distribution one could infer that the partons in the pomeron have rather a hard distribution $f_{q,g/I\!\!P}(z)\sim z(1-z)$ and also indications \cite{Brandt:1992zu} of a superhard component $\sim \delta(1-z)$.

In spite of this discovery, hard diffraction was not fully recognised in the whole particle physics community. It was therefore a surprise to many when diffractive deep inelastic scattering was discovered by ZEUS \cite{ZEUS-DDIS} and H1 \cite{H1-DDIS} at HERA in 1993 through spectacular events having the whole forward detector empty, {\it i.e.}\ a large rapidity gap as opposed to the abundant forward hadronic activity in normal DIS events. A surprisingly large fraction $\sim 10$\% 
of all DIS events were diffractive. Moreover, they showed the same $Q^2$ dependence as normal DIS, demonstrating that they were {\em not} a higher twist process but leading twist. 

The diffractive DIS cross-section can be written\cite{Ingelman:1992qf}
$\frac{d\sigma}{dx\,dQ^2\,dx_{I\!\!P}\,dt}=\frac{2\pi\alpha^2}{xQ^4}
\left( 1+(1-y)^2\right) F_2^{D(4)} $
where fractional energy loss $x_{I\!\!P}$  and four-momentum transfer $t$  from the proton define the diffractive conditions. For most of the data, the leading proton is not observed and hence $t$ is effectively integrated out giving high-precision measurements of the structure function $F_2^{D(3)}(x_{I\!\!P},\beta,Q^2)$ in terms of the model-independent invariant variables $\beta$ and $x_{I\!\!P}$. \footnote{~~$\beta = \frac{-q^2}{2q\cdot (p_p-p_Y)} = \frac{Q^2}{Q^2+M_X^2-t}$ and $x_{I\!\!P} = \frac{q\cdot (p_p-p_Y)}{q\cdot p_p} = \frac{Q^2+M_X^2-t }{Q^2+W^2-M_p^2} = \frac{x}{\beta} $}

In $p\bar{p}$, UA8 has provided more information on diffractive jet production through analyses of such cross-sections\cite{ua8-analyses}. Several different diffractive hard scattering processes have been observed in $p\bar{p}$ at the Tevatron. Events with jets, $W$, $Z$, $b\bar{b}$ or $J/\psi$ have a rapidity gap in about 1\% 
of the cases, {\it i.e.} an order of magnitude smaller relative rate than diffractive DIS at HERA. 

\section{Pomeron approach}\label{sec:pomeron}
The diffractive DIS structure function can be written 
$F_2^{D(4)}(x,Q^2,x_{I\!\!P},t)=f_{I\!\!P /p}(x_{I\!\!P},t) F_2^{I\!\!P}(\beta,Q^2)$, 
in terms of a pomeron flux and a pomeron structure function, where $x_{I\!\!P} \simeq p_{I\!\!P}/p_p$ is interpreted as the momentum fraction of the pomeron in the proton and $\beta \simeq p_{q,g}/p_{I\!\!P}$ is the momentum fraction of the parton in the pomeron. 

Good fits with data can be obtained, provided that also a Reggeon exchange contribution is included. Factoring out the fitted $x_{I\!\!P}$ dependence, one obtains the diffractive structure function $F_2^{D(2)}(\beta,Q^2)$ (or $F_2^{I\!\!P}$). The $Q^2$ dependence is rather weak and thus shows approximate scaling indicating scattering on point-like charges. There is, however, a weak $\log{Q^2}$ dependence which fits well with conventional perturbative QCD evolution. The $\beta$ dependence is quite flat, which can be interpreted as hard parton distributions in the pomeron. This is borne out in a full next-to-leading order QCD fit giving the parton distributions with a dominant gluon component.

Using this model with diffractive parton densities from HERA to calculate diffractive hard processes at the Tevatron, one obtains cross-sections which are about an order of magnitude larger than observed. This problem can be cured by modifications of the model, in particular, by introducing some kind of damping \cite{pomeron-renormalisation} at high energies, such as a pomeron flux `renormalization'. It is, however, not clear whether this is the right way to get a proper understanding.

Thus, there are problems with the pomeron approach. The pomeron flux and the pomeron parton densities do not seem to be universal quantities. They cannot be separately well defined since only their product is experimentally measurable. Moreover, it may be improper to think of the pomeron as `emitted' from the proton, because the soft momentum transfer $t$ at the proton-pomeron vertex imply a long space-time scale such that they move together for an extended time which means that there should be some cross-talk between such strongly interacting objects. In order to investigate these problems, alternative approaches have been investigated where the pomeron is not in the initial state, {\it i.e.}\ not part of the proton wave function but an effect of the QCD dynamics of the scattering process. 

\section{QCD-based approaches}\label{sec:qcd}
\begin{figure}
\hspace*{5mm}
\epsfig{file=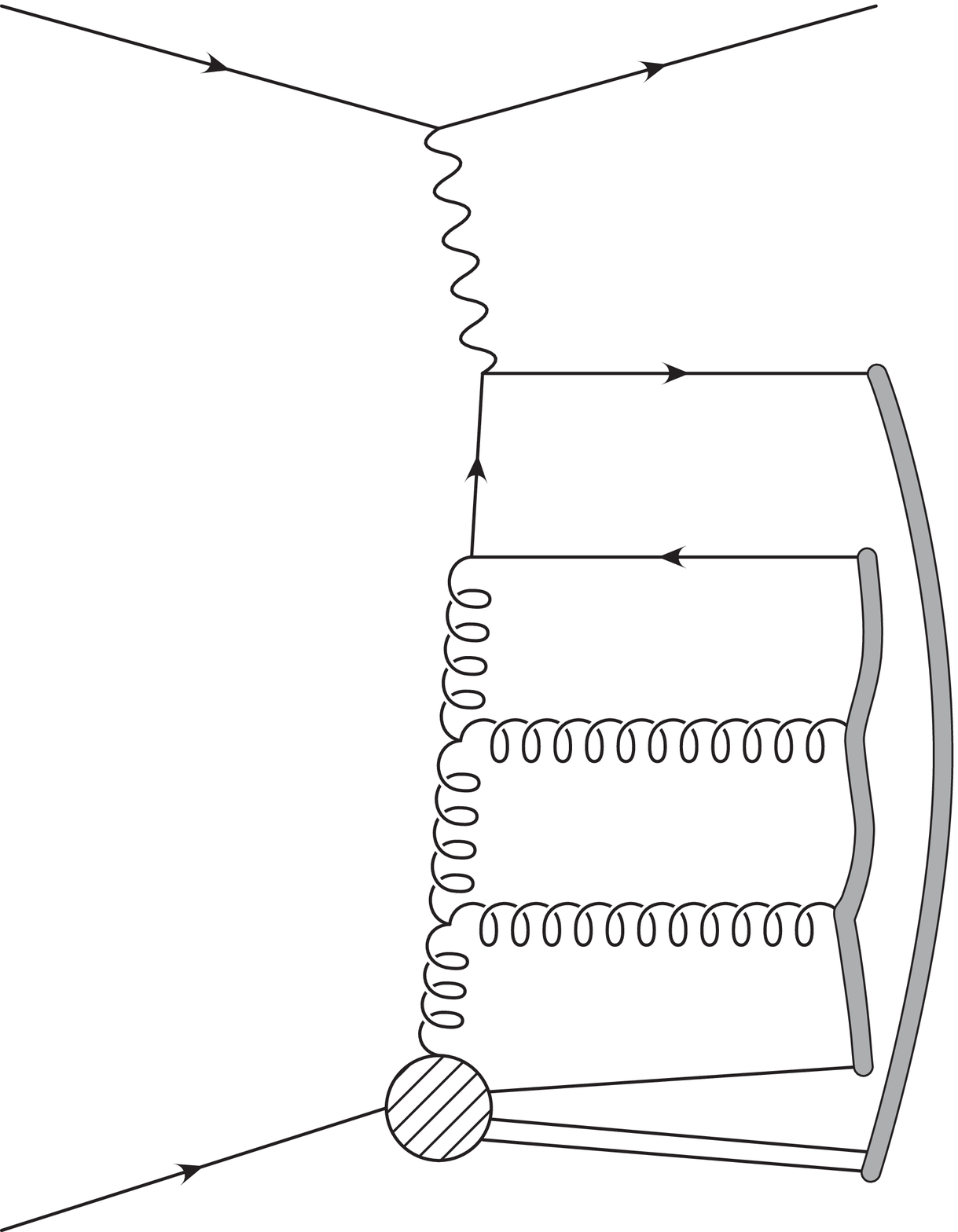,width=0.15\columnwidth}
\epsfig{file=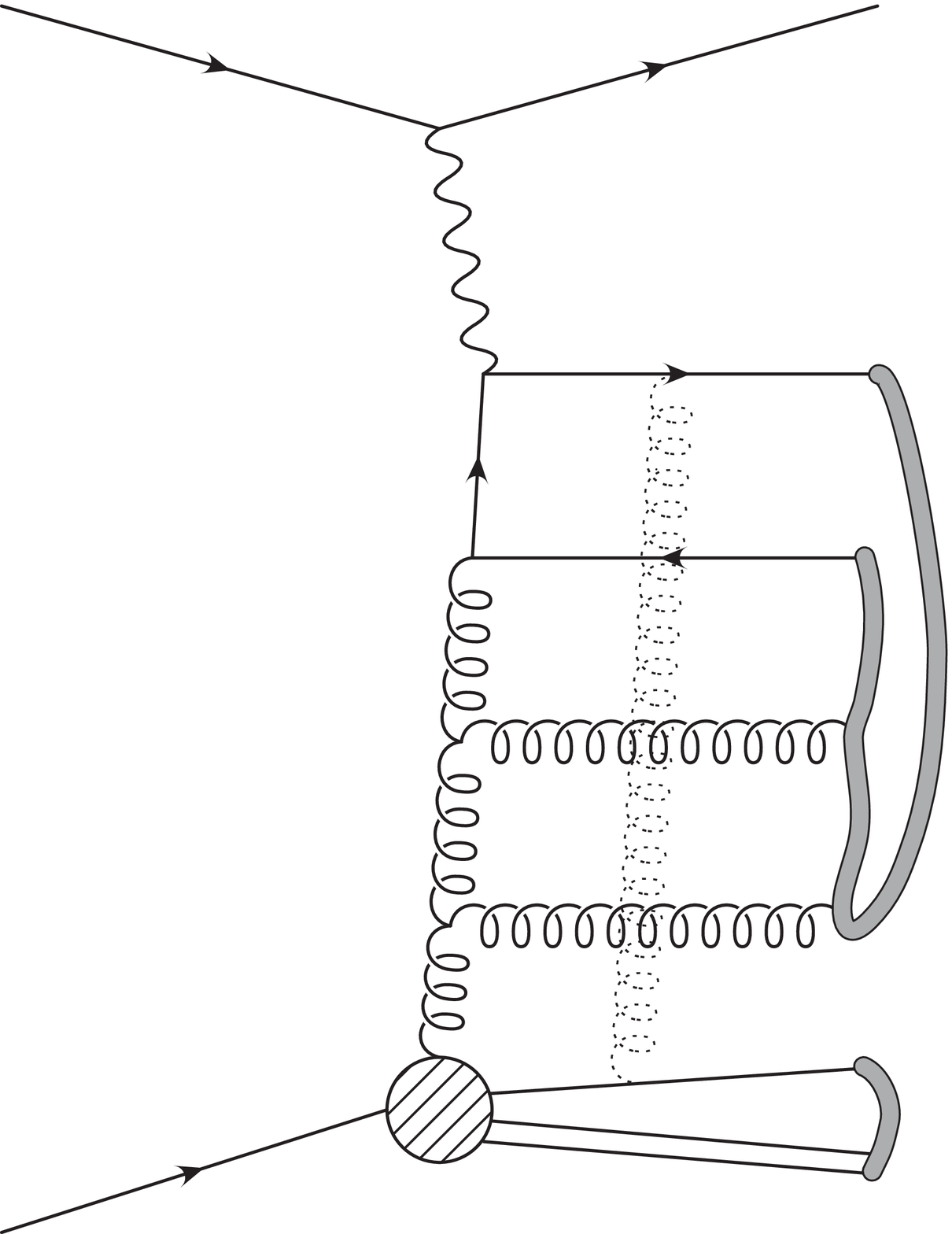,width=0.15\columnwidth}
\vspace*{-40mm}\\  \hspace*{60mm} 
\epsfig{file=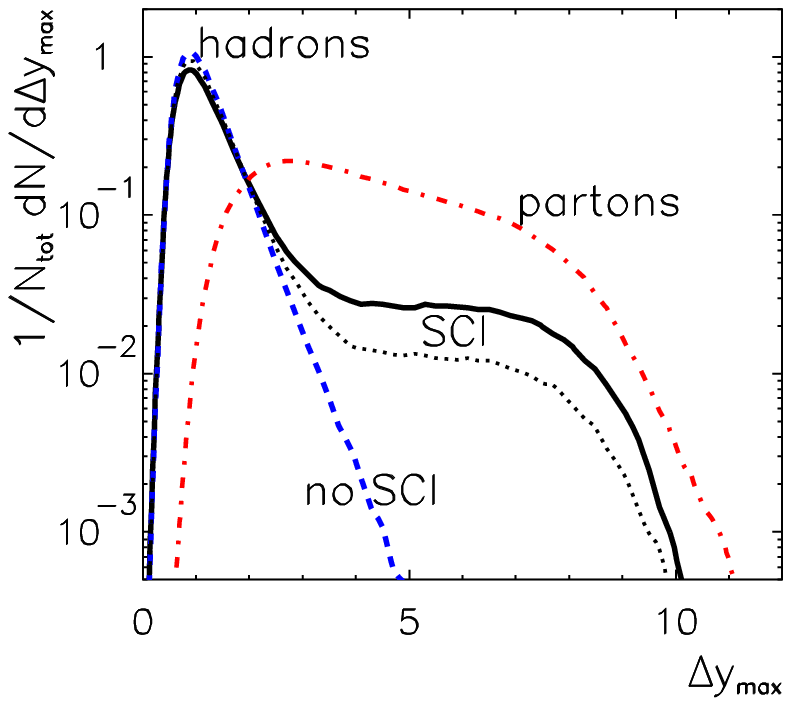,width=0.3\columnwidth}
\hspace*{5mm}
\epsfig{file=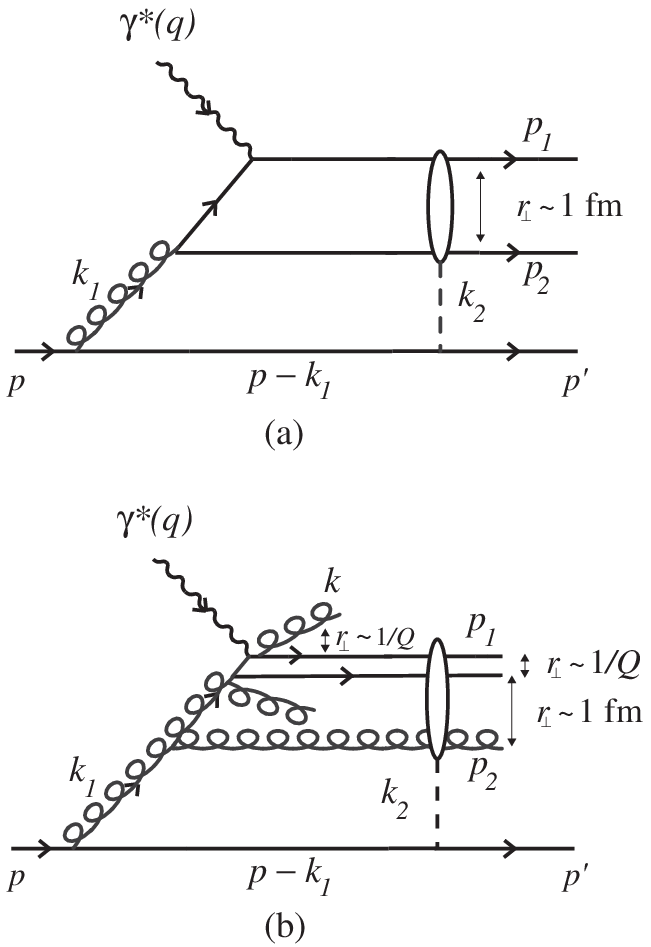, bbllx=0, bblly=0, bburx=186, bbury=274,clip=,width=0.18\columnwidth}
\vspace*{-10mm}\\
\caption{(a,b) Conventional string configuration and after a soft colour-octet exchange (dashed gluon line) between the remnant and the hard scattering system resulting in a rapidity gap. (c) Distribution of the size $\Delta y_{max}$ of the largest rapidity gap in simulated HERA DIS events at parton level (after matrix elements and parton showers) and at hadron level after standard hadronisation (dashed) and when including the SCI model with soft gluon exchange probability $P=0.5$ or $0.1$ (full, dotted).
(d) Rescattering model \protect\cite{Brodsky:2004hi} of a large $q\bar{q}$ or $q\bar{q}g$ system on the target via `instantaneous' longitudinal gluon exchange.}
\label{fig:dis-strings}
\end{figure}

A starting point can here be the standard hard perturbative interactions, since they should not be affected by the soft interactions. On the other hand we know that there should be plenty of soft interactions, below the cut-off $Q^2_0$ for perturbation theory, because $\alpha_s$ is then large giving a large interaction probability ({\it e.g.}\ unity for hadronisation). A simple, but phenomenologically successful attempt in this spirit is the Soft Colour Interaction (SCI) model.\cite{sci} DIS at small $x$ is typically gluon-initiated leading to perturbative parton level processes as illustrated in Fig.~\ref{fig:dis-strings}. The colour order of the perturbative diagram has conventionally been used to define the topology of the resulting non-perturbative colour string fields between the proton remnant and the hard scattering system, such that hadronisation produces hadrons over the full rapidity region. 

We have, however, proper theory only for the hard perturbative part of the event, which is separated from the soft dynamics in both the initial and the final parts by the QCD factorization theorem. This hard part is above a perturbative QCD cut-off $Q^2_0\sim 1 \mathrm{GeV}^2$ with an inverse giving a transverse size which is small compared to the proton diameter. Thus, the hard interactions can be viewed as being embedded in the colour field of the proton and hence one can consider interactions of the outgoing partons with this `background' field. A soft gluon exchange can then rearrange colour so that the hard scattering system becomes a colour singlet and the proton remnant another singlet (Fig.~\ref{fig:dis-strings}) hadronising independently of each other with a rapidity gap in between. The gap can be large because the primary gluon has a small momentum fraction $x_0$, leaving a large momentum fraction $(1-x_0)$ to the remnant which can form a leading proton.

The soft gluon exchange is non-perturbative and hence its probability cannot be calculated. The model, therefore, introduces a single parameter $P$ for the probability of exchanging such a soft gluon between any pair of partons, where one of them should be in the remnant representing the colour background field. Applying this on the partonic state, including remnants, in the Lund Monte Carlo generators {\sc Lepto} for $ep$ and {\sc Pythia} for hadronic interactions, {\it e.g.}\ $p\bar{p}$, leads to variations of the string topologies and thereby different final states after hadronisation. 

One should realize that {\em gap-size is a highly infrared sensitive observable} (Fig.~\ref{fig:dis-strings}c). At the parton level, even after perturbative QCD parton showers, it is quite common to have large gaps. Hadronising the conventional string topology leads to an exponential suppression with the gap-size, {\it i.e.}\ a huge non-perturbative hadronisation effect. Introducing the soft colour interactions causes a drastic effect on the hadron level result, with a gap-size distribution that is not exponentially suppressed but has the plateau characteristic for diffraction. 

\begin{figure}
\hspace*{10mm}
\begin{tabular}{llll}
\multicolumn{4}{c}{$R_{\mathrm{hard}} = \frac{1}{\sigma_{\mathrm{hard}}^{\mathrm{tot}}}
\int_{{x_F}_{\mathrm{min}}}^1 dx_F \, \frac{d\sigma_{\mathrm{hard}}}{dx_F}$}
\vspace*{2mm}\\
\hline
\hline
$R_{\mathrm{hard}} [\%
]$ & \multicolumn{2}{c}{Exp. observed} & SCI \\
\hline
dijets   & {\small CDF} & 0.75 $\pm$ 0.10           & 0.7  \\
W        & {\small CDF} & 1.15 $\pm$ 0.55           & 1.2  \\
W        & {\small D\O} & 1.08 $^{+0.21}_{-0.19}$   & 1.2  \\
$b\bar{b}$   & {\small CDF} & 0.62 $\pm$ 0.25           & 0.7  \\
Z        & {\small D\O} & 1.44 $^{+0.62}_{-0.54}$   & 1.0$^\star$ \\
$J/\psi$ & {\small CDF} & 1.45 $\pm$ 0.25           & 1.4$^\star$ \\
\hline
\hline
\end{tabular}
\vspace*{-50mm}\\ \hspace*{90mm}
\epsfig{file=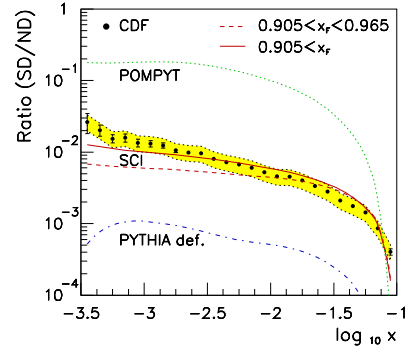,width=0.35\columnwidth}
\vspace*{-10mm}\\
\caption{(a) Tevatron data on the ratio of diffractive hard processes to all such hard events, compared to the SCI model. (b) Ratio of diffractive to non-diffractive dijet events versus momentum fraction $x$ of the interacting parton in $\bar{p}$, with CDF data compared to the {\sc Pompyt} pomeron model, default {\sc Pythia} and the SCI model.\protect\cite{Enberg:2001vq}}
\label{fig:dijets-x}
\end{figure}

Selecting the gap events in the Monte Carlo one can extract the diffractive structure function and the model (choosing $P\approx 0.5$) describes quite well \cite{sci} the main features of $F_2^{D(3)}(x_{I\!\!P},\beta,Q^2)$ observed at HERA. Using exactly the same model applied to $p\bar{p}$ at the Tevatron, one obtains the correct overall rates of diffractive hard processes as observed at the Tevatron (Fig.~\ref{fig:dijets-x}).  Differential distributions are also reproduced \cite{Enberg:2001vq} as exemplified in Fig.~\ref{fig:dijets-x}b, which also demonstrates that the pomeron model is far above the data and {\sc Pythia} without the SCI mechanism is far below. 

The phenomenological success of the SCI model indicates that it captures the most essential QCD dynamics responsible for gap formation. It is therefore interesting that recent theoretical developments provide a basis for this model. The QCD factorization theorem separates the hard and soft dynamics and is the basis for the definition of the parton density functions 
$f_{q/N}  \sim \int dx^- e^{-i x_B p^+ x^-/2} \langle \,N(p)\,|\, \bar\psi(x^-) 
\gamma^+\, W[x^-;0] \, \psi(0)\,|\,N(p)\,\rangle_{x^+=0}$. 
Here, the nucleon state sandwiches an operator including the Wilson line
$W[x^-;0] = {\rm P}\exp\left[ig\int_0^{x^-} dw^- A_a^+(0,w^-,0_\perp) t_a \right]$ which is a path-ordered exponential of gluon fields. The physical interpretation becomes transparent if one expands the exponential giving \cite{Brodsky:2002ue}
$$
W[x^-;0] \sim 1 + g\int \frac{dk_1^+}{2\pi} \frac{\tilde A^+(k_1^+)}{k_1^+-i\varepsilon} + g^2 \int \frac{dk_1^+ dk_2^+}{(2\pi)^2}
\frac{\tilde A^+(k_1^+)\tilde A^+(k_2^+)}{(k_1^+ +k_2^+-i\varepsilon)
(k_2^+-i\varepsilon)} + \ldots  
$$
with terms of different orders in the strong coupling $g$. The first term is the scattered `bare' quark and the following terms corresponds to rescattering on the target colour field via 1,2\ldots gluons. This rescattering \cite{Brodsky:2002ue} has leading twist contributions for longitudinally polarised gluons, which are instantaneous in light-front time $x^+=t+z$ and occurs within Ioffe coherence length $\sim 1/m_px_{Bj}$ of the hard DIS interaction. 

This implies a rescattering of the scattered quark with the spectator system in DIS. Although one can choose a gauge such that the scattered quark has no rescatterings, one can not `gauge away' all rescatterings with the spectator system.\cite{Brodsky:2002ue} The sum of the couplings to the $q\bar{q}$-system in Fig.~\ref{fig:dis-strings}d gives the same result in any gauge and is equivalent to the colour dipole model in the target rest system (discussed below). Thus, there will always be such rescatterings and their effects are absorbed in the parton density functions obtained by fitting inclusive DIS data.

This has recently been used \cite{Brodsky:2004hi} as a basis for the SCI model. A gluon from the proton splits into a $q\bar{q}$ pair that the photon couples to (Fig.~\ref{fig:dis-strings}d). Both the gluon and its splitting are mostly soft since this has higher probability ($g(x)$ and $\alpha_s$). The produced $q\bar{q}$ pair is therefore typically a large colour dipole that even a soft rescattering gluon can resolve and therefore interact with. The discussed instantaneous gluon exchange can then modify the colour topology before the string-fields are formed such that colour singlet systems separated in rapidity arise producing a gap in the final state after hadronisation. Similarly, the initial gluon may also split softly into a gluon pair (Fig.~\ref{fig:dis-strings}d) followed by perturbative $g\to q\bar{q}$ giving a small $q\bar{q}$ pair. Soft rescattering gluons can then not resolve the $q\bar{q}$, but can interact with the large-size $q\bar{q}$--$g$ colour octet dipole and turn that into a colour singlet system separated from the target remnant system that is also in a colour singlet state. Higher order perturbative emissions do not destroy the gap, since it occurs in the rapidity region of the hard system and not in the gap region. 

The rescattering produces on-shell intermediate states having imaginary amplitudes,\cite{Brodsky:2002ue} which is a characteristic feature of diffraction. This theoretical framework implies the same $Q^2,x,W$ dependencies in both diffractive and non-diffractive DIS, in accordance with HERA data. 

Another approach, which has some similarities, is the so-called semi-classical approach.\cite{buchmueller-hebecker} The analysis is here made in the proton rest frame where the incoming photon fluctuates into a $q\bar{q}$ pair or a $q\bar{q}$--$g$ system that traverses the proton. The soft interactions of these colour dipoles with the non-perturbative colour field of the proton is estimated using Wilson lines describing the interaction of the energetic partons with the soft colour field of the proton. The colour singlet exchange contribution to this process has been derived and shown to give leading twist diffraction when the dipole is large. This corresponds to a dipole having one soft parton (as in the aligned jet model), which is dominantly the gluon. One is thus testing the large distances in the proton colour field. This soft field cannot be calculated from first principles and is therefore modelled involving parameters fitted to data. This theoretical approach is quite successful in describing the HERA data on $F_2^{D(3)}(x_{I\!\!P};\beta,Q^2)$. 

Yet another approach starts from perturbative QCD and attempts to describe diffractive DIS as a two-gluon exchange.\cite{Bartels:1998ea} Here, the photon fluctuates into a $q\bar{q}$ or a $q\bar{q}$--$g$ colour dipole, which couples to the two gluons as calculated in perturbative QCD giving essentially the $\beta$ and $Q^2$ dependencies of the diffractive structure function 
$x_{I\!\!P}F_2^{D(3)}= F_{q\bar{q}}^T + F_{q\bar{q}g}^T + F_{q\bar{q}}^L$, 
with contributions of $q\bar{q}$ and $q\bar{q}g$ colour dipoles from  photons with transverse ($T$) and longitudinal ($L$) polarization. The connection of the exchanged gluons with the proton cannot, however, be calculated perturbatively. This soft dynamics is introduced through a parameterisation where one fits the $x_{I\!\!P}$ dependence, which introduces parameters for the absolute normalization. The result \cite{Chekanov:2005vv} is a quite good fit to the data and the different contributions from the two dipoles and photon polarizations can be investigated and provide interesting information on the QCD dynamics described by this approach. 

\section{Gaps between jets and BFKL}\label{sec:gaps-bfkl}
\begin{figure}
\hspace*{25mm}
\epsfig{file=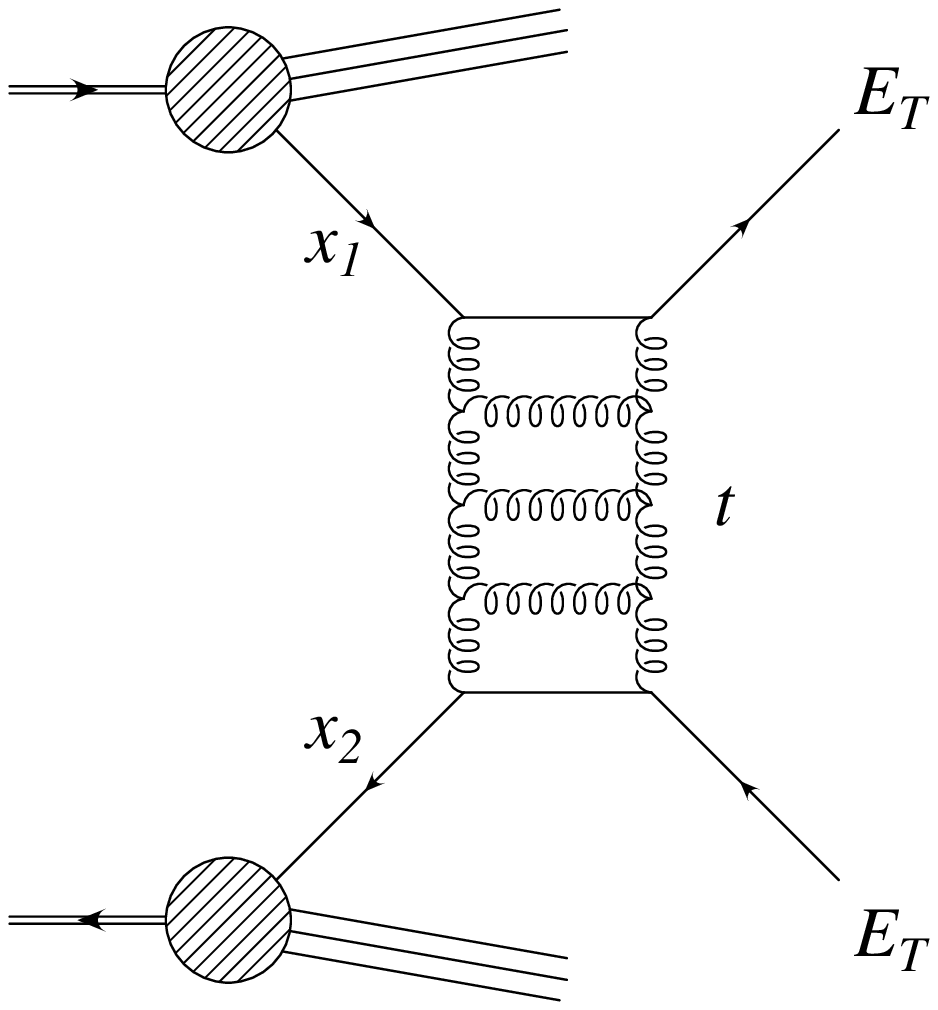,width=0.18\columnwidth}
\vspace*{-40mm}\\ \hspace*{70mm}
\epsfig{file=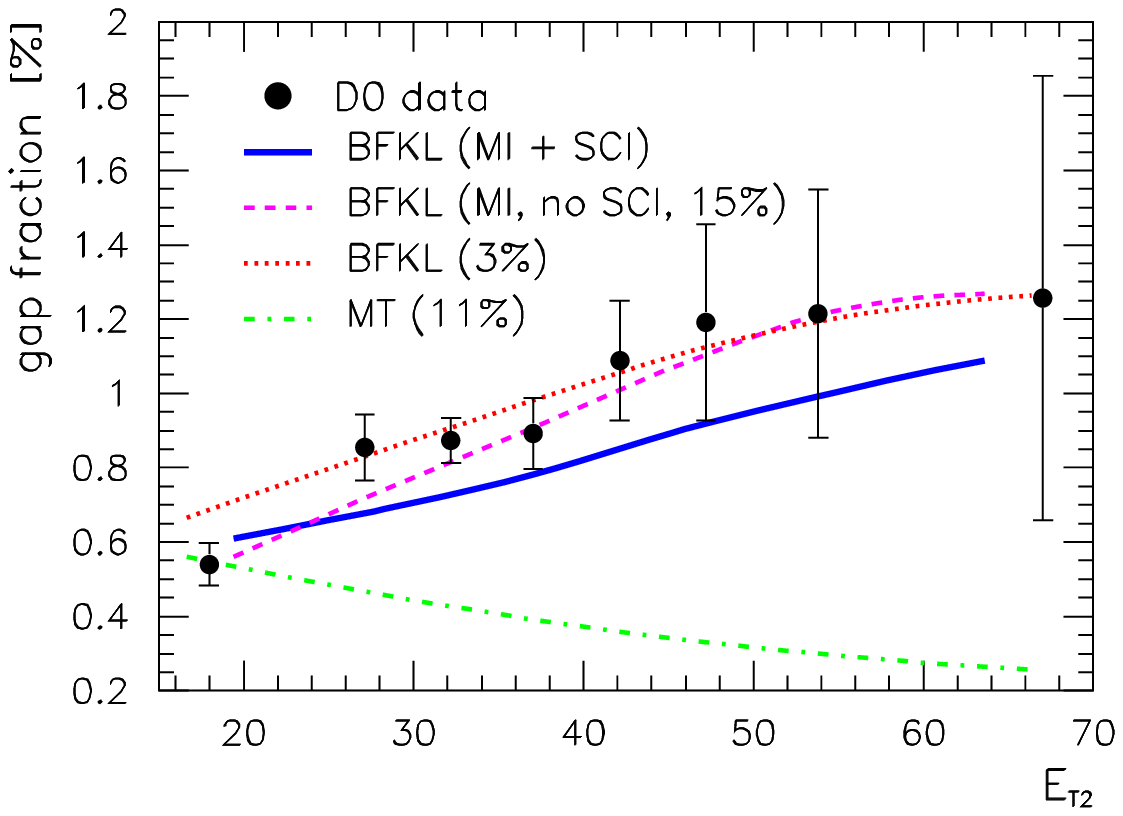,width=0.35\columnwidth}
\vspace*{-10mm}\\
\caption{(a) Colour singlet exchange by a BFKL gluon ladder giving a rapidity gap between two jets. (b) Fraction of events with a rapidity gap between the jets versus jet-$E_T$, where D0 data is compared to the BFKL colour singlet exchange mechanism\protect\cite{Enberg:2001ev} with underlying event treatments: simple 3\% 
gap survival probability, {\sc Pythia}'s multiple interactions (MI) and hadronisation requiring a 15\% 
gap survival probability, MI plus SCI model and hadronisation (no gap survival factor). Also, asymptotic Mueller-Tang (MT) with an 11\%
gap survival probability.}
\label{fig:BFKL-ladder}
\end{figure}

In the processes discussed so far, the hard scale has not involved the gap itself since the leading proton has only been subject to a soft momentum transfer across the gap. A new milestone was therefore the observation \cite{D0-CDF-gap-jet-gap} at the Tevatron of events with a gap between two high-$p_\perp$ jets. This means that there is a large momentum transfer across the gap and perturbative QCD should therefore be applicable to understand the process. This is indeed possible by considering elastic parton-parton scattering via hard colour singlet exchange in terms of two gluons as illustrated in Fig.~\ref{fig:BFKL-ladder}. In the high energy limit $s/|t| \gg 1$, where the parton cms energy is much larger than the momentum transfer, the amplitude is dominated by terms $\sim [\alpha_s \, \ln (s/|t|)]^n$ where the smallness of $\alpha_s$ is compensated by the large logarithm. These terms must therefore be resummed leading to the famous BFKL equation describing the exchange of a whole gluon ladder.

This somewhat complicated equation has been solved numerically.\cite{Enberg:2001ev} This gave the matrix elements for an effective $2\to 2$ parton scattering process, which was implemented in the Lund Monte Carlo {\sc Pythia} such that parton showers and hadronisation could be added to generate complete events. This reproduces the data, both in shape and absolute normalization, which is not at all trivial (Fig.~\ref{fig:BFKL-ladder}). The included non-leading corrections are needed since the asymptotic Mueller-Tang result has the wrong $E_T$ dependence. A free gap survival probability parameter, which in other models is introduced to get the correct overall normalization, is not needed in this approach. Amazingly, the correct gap rate results from the complete model, including the above discussed soft colour interaction model. 

\section{Conclusions}\label{sec:conclusions}
After 20 years, hard diffraction has developed into an important research field, with a lot of high-quality data from $p\bar{p}$ and $ep$. Theory and models have provided working phenomenological descriptions, but we do not have solid theory yet. 

In the new QCD-based models, the pomeron is not part of the proton wave function, but diffraction is an effect of the scattering process. Models based on interactions with a colour background field provide an interesting approach which avoids conceptual problems of pomeron-based models, such as the pomeron flux, and provide a basis for a common theoretical framework for all final states, diffractive gap events as well as non-diffractive events. 

Finally, the new process of gaps between jets provides strong evidence for the BFKL dynamics as predicted since long by QCD, but so far hard to establish experimentally. 



\section*{References}
\begin{thebibliography}{99}

\bibitem{Ingelman:1985ec}
G.~Ingelman,
p.\ 135 in `Elastic and Diffractive Scattering', Eds.\ B.\ Nicolescu, J.\ Tran Than Van, Editions Fronti\`eres 1985

\bibitem{Ingelman:1984ns}
G.~Ingelman and P.E.~Schlein,
Phys.\ Lett.\ B {\bf 152}, 256 (1985)

\bibitem{Bonino:1988ae}
R.~Bonino {\it et al.}  [UA8 Collaboration],
Phys.\ Lett.\ B {\bf 211}, 239 (1988)

\bibitem{Brandt:1992zu}
A.~Brandt {\it et al.}  [UA8 Collaboration],
Phys.\ Lett.\ B {\bf 297}, 417 (1992)

\bibitem{ZEUS-DDIS}
M.~Derrick {\it et al.}  [ZEUS],
Phys.\ Lett.\ B {\bf 315}, 481 (1993); 
Z.\ Phys.\ C {\bf 68}, 569 (1995)

\bibitem{H1-DDIS}
T.~Ahmed {\it et al.}  [H1],
Nucl.\ Phys.\ B {\bf 429}, 477 (1994); 
Phys.\ Lett.\ B {\bf 348}, 681 (1995)

\bibitem{Ingelman:1992qf}
G.~Ingelman and K.~Prytz,
Z.\ Phys.\ C {\bf 58}, 285 (1993)

\bibitem{ua8-analyses}
A.~Brandt {\it et al.}  [UA8],
Phys.\ Lett.\ B {\bf 421}, 395 (1998); 
Eur.\ Phys.\ J.\ C {\bf 25}, 361 (2002)

\bibitem{pomeron-renormalisation}
K.~Goulianos and J.~Montanha,
Phys.\ Rev.\ D {\bf 59}, 114017 (1999)
\\
S.~Erhan and P.~E.~Schlein,
Phys.\ Lett.\ B {\bf 427}, 389 (1998)

\bibitem{sci}
A.~Edin, G.~Ingelman and J.~Rathsman,
Phys.\ Lett.\ B {\bf 366}, 371 (1996); 
Z.\ Phys.\ C {\bf 75}, 57 (1997); 
DESY-PROC-1999-02, arXiv:hep-ph/9912539

\bibitem{Enberg:2001vq}
R.~Enberg, G.~Ingelman and N.~Timneanu,
Phys.\ Rev.\ D {\bf 64}, 114015 (2001)

\bibitem{Brodsky:2002ue}
S.~J.~Brodsky {\it et al.} 
Phys.\ Rev.\ D {\bf 65}, 114025 (2002)

\bibitem{Brodsky:2004hi}
S.~J.~Brodsky, R.~Enberg, P.~Hoyer and G.~Ingelman,
Phys.\ Rev.\ D {\bf 71}, 074020 (2005)

\bibitem{buchmueller-hebecker}
A.~Hebecker,
Phys.\ Rept.\  {\bf 331}, 1 (2000)

\bibitem{Bartels:1998ea}
J.~Bartels, J.~R.~Ellis, H.~Kowalski and M.~Wusthoff,
Eur.\ Phys.\ J.\ C {\bf 7}, 443 (1999)
 
\bibitem{Chekanov:2005vv}
S.~Chekanov {\it et al.}  [ZEUS Collaboration],
Nucl.\ Phys.\ B {\bf 713}, 3 (2005)

\bibitem{D0-CDF-gap-jet-gap} 
F.~Abe {\it et al.}  [CDF Collaboration],
Phys.\ Rev.\ Lett.\  {\bf 80}, 1156 (1998)
\\
B.~Abbott {\it et al.}  [D0 Collaboration],
Phys.\ Lett.\ B {\bf 440}, 189 (1998)

\bibitem{Enberg:2001ev}
R.~Enberg, G.~Ingelman and L.~Motyka,
Phys.\ Lett.\ B {\bf 524}, 273 (2002)

\end{thebibliography}
\end{document}